\documentclass{iau_eggl}
\usepackage{graphicx}
\usepackage{natbib}

\usepackage{amssymb}
\usepackage{amsmath}
\usepackage{amsfonts}

\usepackage{epsf,pstricks}
\usepackage{epsfig}
\usepackage{colordvi}
\usepackage{color}

\definecolor{darkgreen}{rgb}{0,0.8,0}
\definecolor{darkyellow}{rgb}{0.8,0.8,0}

\title[Dynamics and Habitability] 
{Dynamics and Habitability in Binary Star Systems}

\author[S. Eggl, N. Georgakarakos \& E. Pilat-Lohinger]   
{Siegfried Eggl$^1$,
  Nikolaos Georgakarakos$^2$
 \and Elke Pilat-Lohinger$^3$}

\affiliation{$^1$IMCCE, Observatoire de Paris, 77 Avenue Denfert-Rochereau,
F-75014, Paris, France \\ email: {\tt siegfried.eggl@imcce.fr} \\[\affilskip]
$^2$ Higher Technological Educational Institute of Central Macedonia, Terma Magnesias, Serres 62124, Greece \\email: {\tt georgakarakos@hotmail.com}\\[\affilskip]
$^3$ University of Graz, Institute of Physics, IGAM, Universit\"atsplatz 5, 8010 Graz, Austria \\email: {\tt elke.pilat-lohinger@univie.ac.at}
}

\pubyear{2014}
\volume{310}  
\pagerange{xxx--xxx}
\jname{Complex Planetary Systems}
\editors{Z. Knezevic \& A. Lema\^itre}
\begin{document}

\maketitle

\begin{abstract}
Determining planetary habitability is a complex matter, as the interplay between a planet's physical and atmospheric properties with  
stellar insolation has to be studied in a self consistent manner. Standardized atmospheric models for Earth-like planets exist and are commonly accepted as a reference 
for estimates of Habitable Zones.
In order to define Habitable Zone boundaries, circular orbital configurations around main sequence stars are generally assumed. 
In gravitationally interacting multibody systems, such as double stars, however, 
 planetary orbits are forcibly becoming non circular with time. Especially in binary star systems even relatively small 
 changes in a planet's orbit can have a large impact on habitability. Hence,  
 we argue that a minimum model for calculating Habitable Zones in binary star systems has to include dynamical interactions. 

\end{abstract}
\firstsection 
\section{Introduction}
One of the most intriguing aspects of exoplanet science is the prospect of finding worlds around other stars that might be capable of hosting life as we know it.
This translates into a search for terrestrial planets in the so-called Habitable Zones (HZs), i.e. circumstellar regions where
an Earth-analog is capable of sustaining liquid water on its surface.
Finding planets in such zones does not automatically mean that they are habitable, though.   
The complex interplay between insolation, atmosphere, lithosphere and hydrosphere necessitates a determination of most planetary, 
stellar and orbital properties in order to judge potential habitability. 
Hence, each newly discovered candidate has to be assessed on an individual basis. 
This, of course, raises questions on the purpose and usefulness of the HZ concept.
Yet, we believe that providing planet hunters with guidelines on where to look for potentially habitable worlds remains 
important, as long as observational resources are limited. This implies that predicted HZ borders have to be translatable into 
detectability domains for the respective exoplanet detection methods.
HZ estimates should, therefore, contain reasonably detailed assumptions on the physical processes underlying planetary habitability.
We argue, for instance, that dynamically induced changes in a planet's orbit can substantially change habitability conditions.
If quantifiable, they should not be neglected in HZ calculations. 
\firstsection
\section{Insolation and Orbital Dynamics}
So far, considerable efforts have been spent on investigating climatic stability of Earth-like planets leading to 
largely accepted estimates on the amount and spectral distribution of insolation that can render a world uninhabitable \citep{kasting-et-al-1993, selsis-et-al-2007,kopparapu-et-al-2013erratum}. 
Recent so-called "effective insolation" limits can be found, for instance, in \citet{kopparapu-et-al-2014}.
Naturally, globally averaged climate models such as used by the previously named authors cannot account for all effects \citep[e.g.][]{wang-et-al-2014,leconte-et-al-2013}, 
but they can be considered a reasonable first approximation for
Earth-like planets that are far from the tidal-lock or tidal-heating regime.
Alas, HZ borders can be calculated by solving the simultaneous equations:
\begin{equation}
S/I \leq 1 \; \text{and} \; S/O \geq 1,\quad \text{leading to} \quad  r_{\{I,O\}}= \left(\frac{L_\star}{4\pi S_c \{I,O\}}\right)^{1/2} \label{eq:1},
\end{equation}
where $S$ is the insolation at the top of the atmosphere, $L_\star$ is the luminosity of the star, $I=S_{eff}^{inner}$ 
and $O=S_{eff}^{outer}$ denote the effective insolation limits for the inner and outer edge of the HZ, and $S_c=1367\,[W/m^2]$ refers to the Solar constant.
For the Sun and an Earth-like planet \cite{kopparapu-et-al-2014} predict $I\sim1.11$ and $O\sim0.36$, resulting in the following HZ limits expressed as distances $r$ from the host star:
$r_I\sim0.95$ au and $r_O\sim1.67$ au.
Equation~(\ref{eq:1}) describes a spherical shell around the host star where planets can populate orbits that respect the insolation limits necessary for habitability. 
Yet, what if the planet leaves this shell from time to time?
This may happen when the planet's orbit is elliptic. Then, the top of the atmosphere 
insolation can vary considerably with time, see Figure~\ref{fig1}. 
\citet{williams-pollard-2002} could show that planets with Earth-like oceans can compensate climatic extremes due to excursion outside the 
classical HZs (henceforth CHZ, not to be confused with the Continuous Habitable Zone), as long as the average insolation stays 
within habitable limits. Later studies indicate, however, that the role of planetary eccentricity and its influence on insolation cannot easily be
discarded \citep[e.g.][]{spiegel-et-al-2010,dressing-et-al-2010}.
We, thus, propose a way to include available information on the variability of planetary insolation directly into habitability considerations.
\firstsection
\section{A Minimum Dynamical Model for HZs}
\begin{figure}
 \begin{tabular}{ccc}
\includegraphics[scale=0.18, angle=0]{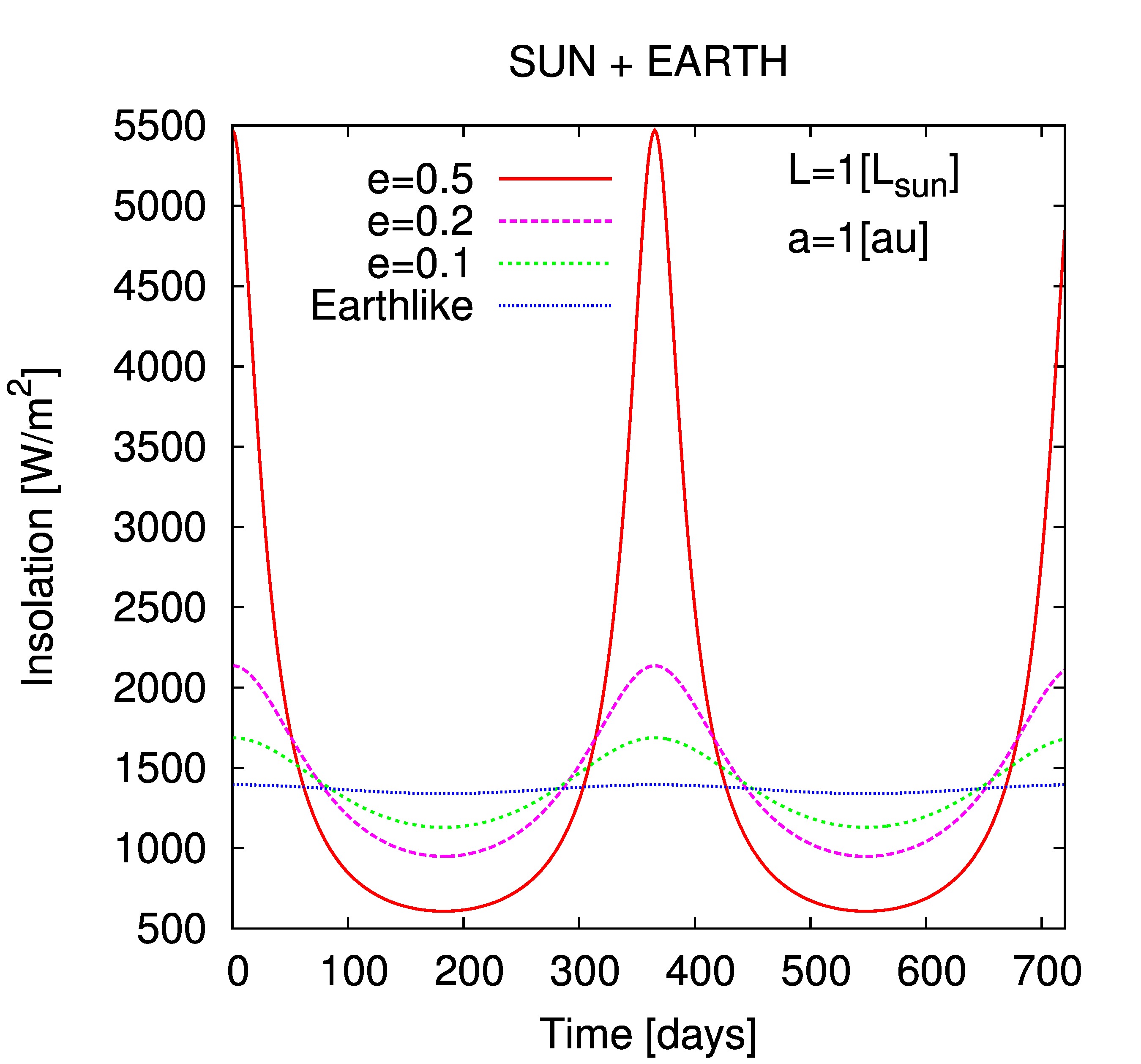} &
\includegraphics[scale=0.18, angle=0]{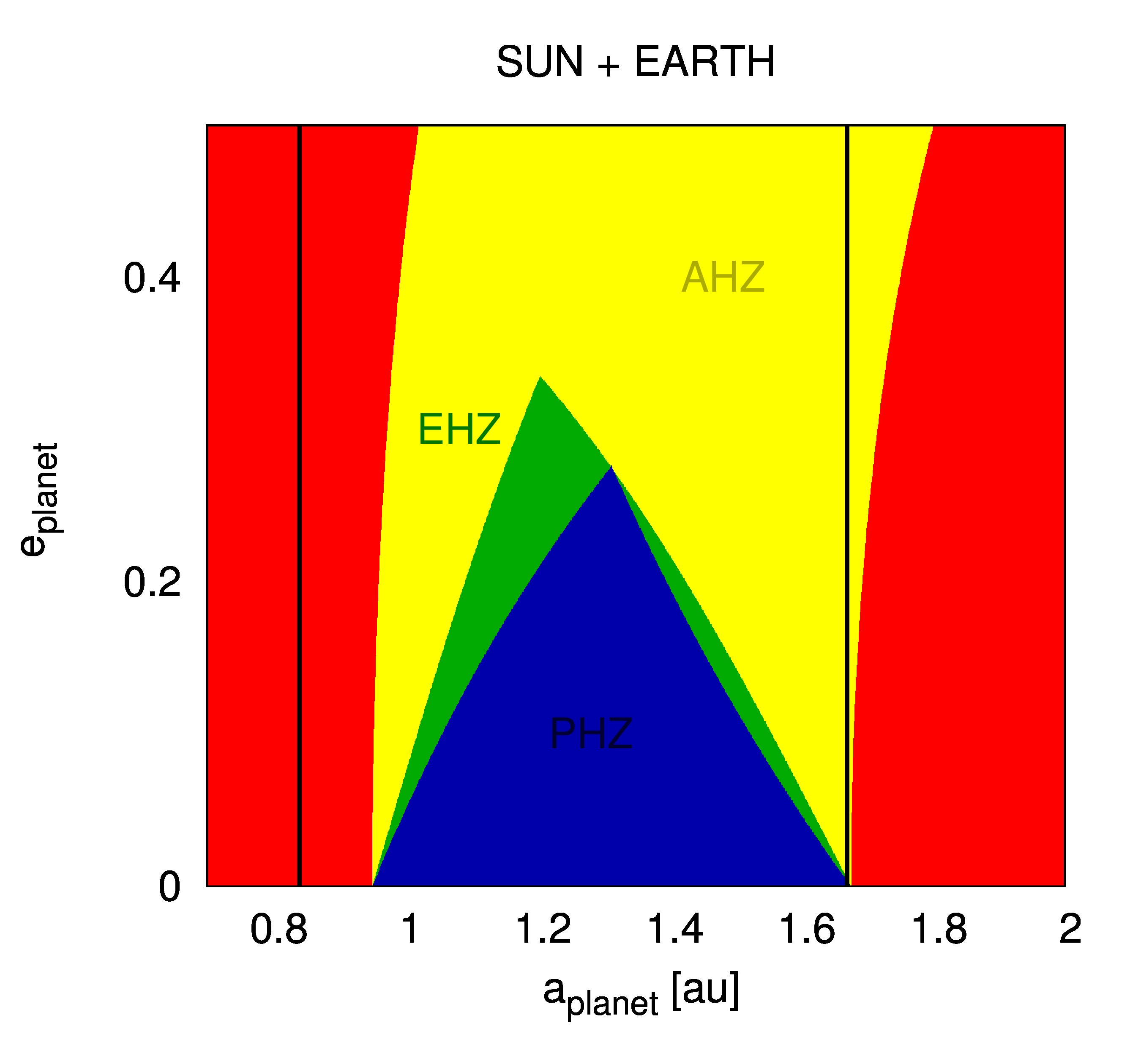} &
 \includegraphics[scale=0.08, angle=0]{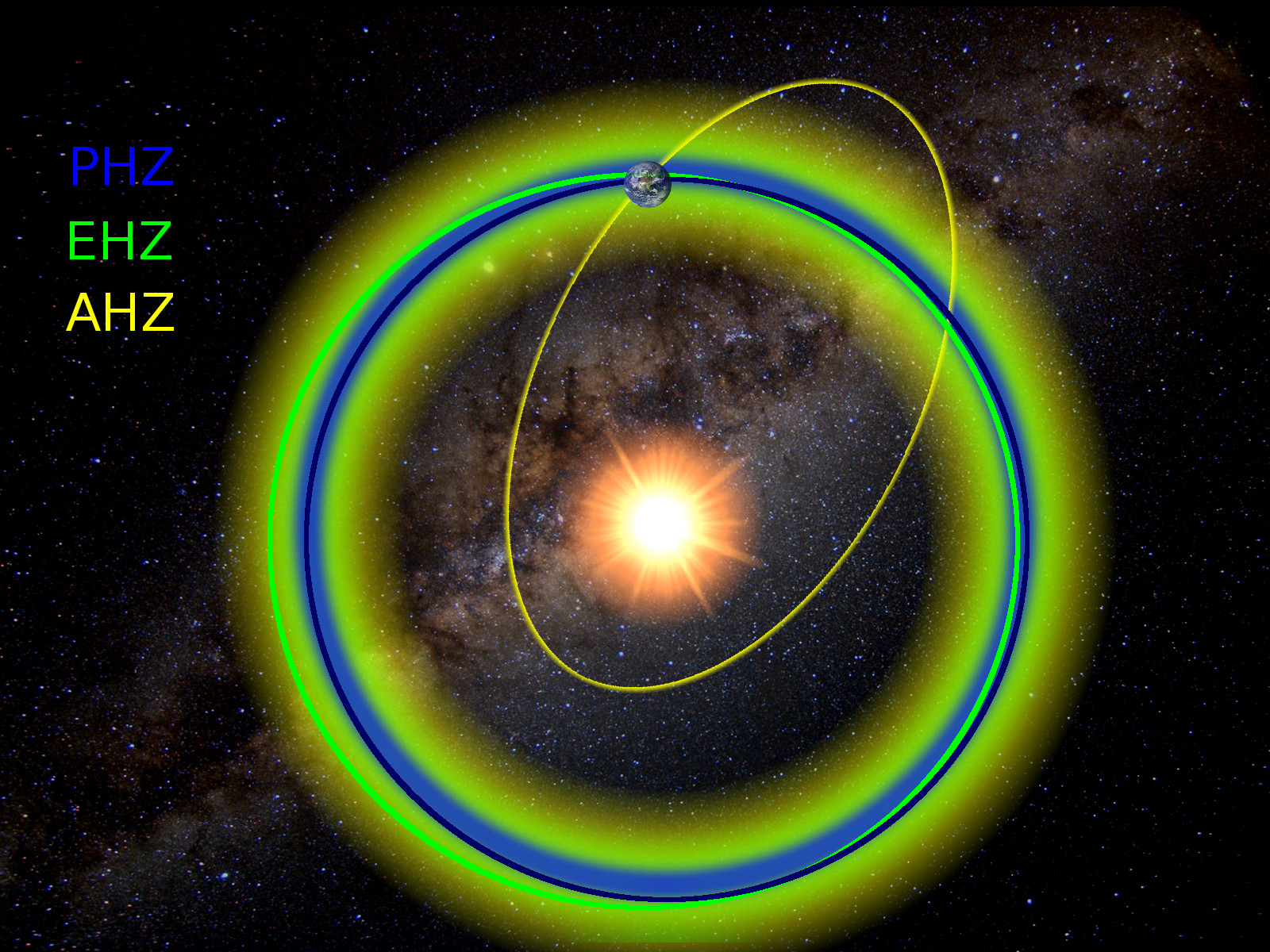}
\end{tabular}
\caption{
\textit{left:} Insolation variability in the two body problem Sun + Earth for mutual orbits with various eccentricities.
\textit{mid:} A habitability map (HM) showing different kinds of Habitable Zones for an Earth-twin orbiting a Sun-like star on constant orbits with various semi-major axes and eccentricities.
The different shades denote: PHZ, EHZ, AHZ and non-habitable regions.
The black vertical lines correspond to the standard CHZ borders after \citet{kasting-et-al-1993} while the shaded HZs have been calculated using
the values by \citet{kopparapu-et-al-2014}. One can see that the PHZ and EHZ
decrease rapidly with mounting planetary eccentricities while the AHZ remains practically constant up to 
high $e_{planet}$ values. The difference between Kasting's and Kopparapu's HZ borders ($e_{planet}=0$) also shows the importance of a classification scheme that can adapt to 
alterations in the climate model. \textit{right:} A concept of the proposed HZ scheme that includes information on the variability of planetary insolation.
\label{fig1}}
\end{figure}
\noindent 
Our aim is to acquire self consistent HZ estimates with a minimum of complexity that still retain the relevant physics.  
To this end we use the effective insolation limits of a globally averaged atmosphere model, where variations in the planet's rotation state can be ignored, as long as its atmospheric properties do not change radically. 
Alterations in the activity of the star could be included in theory, but they are difficult to model and shall be neglected for the moment.
In such a scenario, changes that are related to the planet's orbit can be considered the dominating cause for insolation variability.
In order to include such information into our HZ model, we follow \citet{eggl-et-al-2012} and introduce the concepts of  
the Permanently (PHZ), Extended (EHZ) and the
Averaged Habitable Zones (AHZ), see Figure~\ref{fig1}.
The PHZ is the region where a planet \textit{always} stays within the effective insolation limits ($I$, $O$), 
i.e. \mbox{$S/I \leq 1\; \text{and} \;  S/O \geq 1$}. This is the "classical" definition of a HZ. 
The EHZ allows parts of the planetary orbit to lie beyond the classical HZ:
\mbox{$\left \langle S/I\right \rangle_t +\sigma_I \leq 1 \; \text{and} \;  \left \langle S/O\right \rangle_t -\sigma_O \geq 1$},
where  $\left \langle S \right \rangle_t$ denotes the time-averaged effective insolation and $\sigma^2$ the effective insolation variance.   
High planetary eccentricity may not be prohibitive for habitability since the atmosphere
can act as a buffer. Consequently, the AHZ encompasses all configurations that support the planet's time-averaged effective 
insolation to be within classical limits, e.g. also orbits with very high eccentricities:
\mbox{$ \left \langle S/I \right \rangle_t \leq 1 \; \text{and} \;  \left \langle S/O \right \rangle_t  \geq 1$}.
Figure~\ref{fig1} shows a realization of this classification scheme for an Earth-like planet orbiting a Sun-like star.
\firstsection
\section{Habitable Zones in Binary Star Systems}
\begin{figure}
\begin{center}
\begin{tabular}{cc}
circumstellar planet (S-type) & circumbinary planet (P-type) \\
\includegraphics[scale=0.26, angle=0]{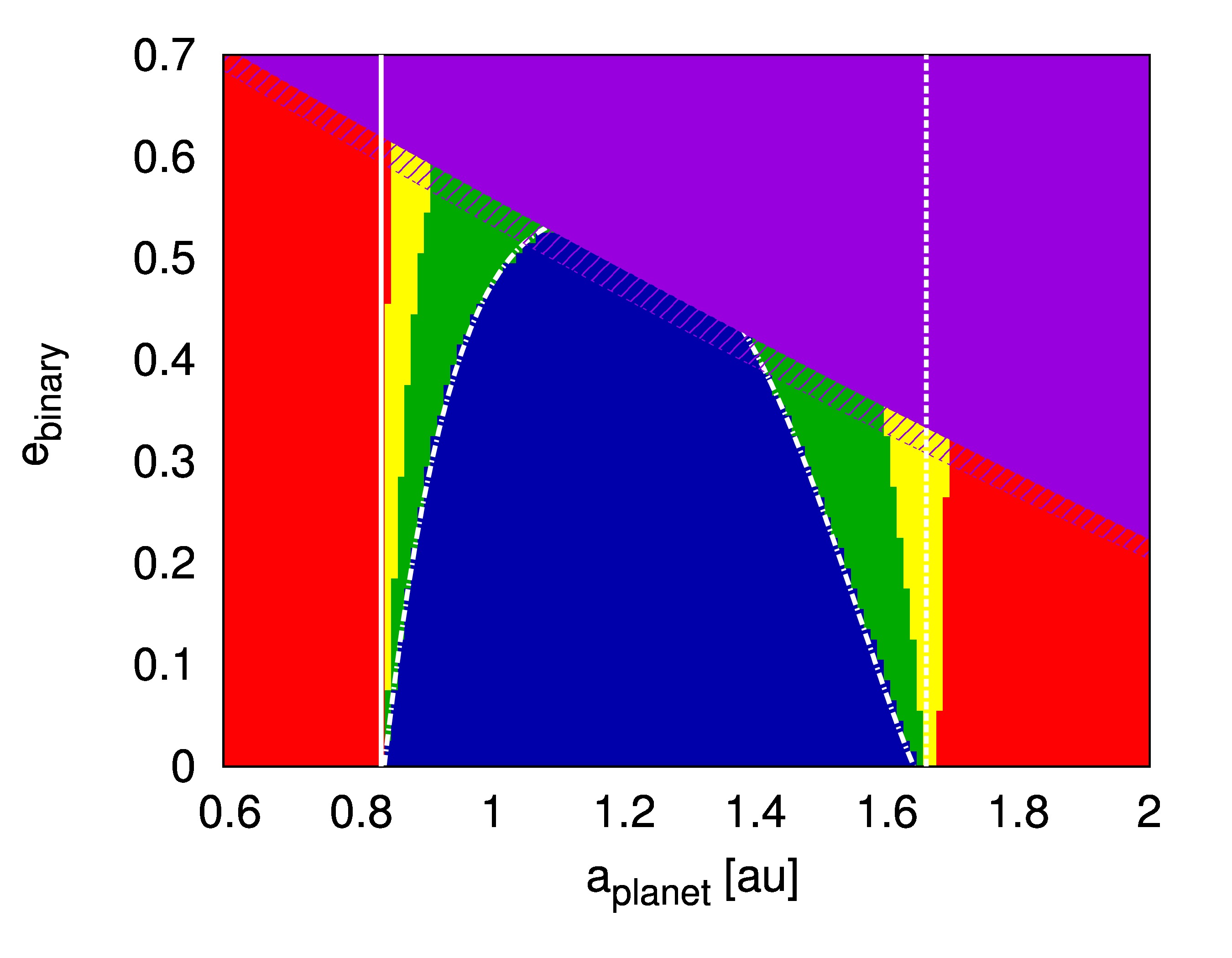} &
\includegraphics[scale=0.26, angle=0]{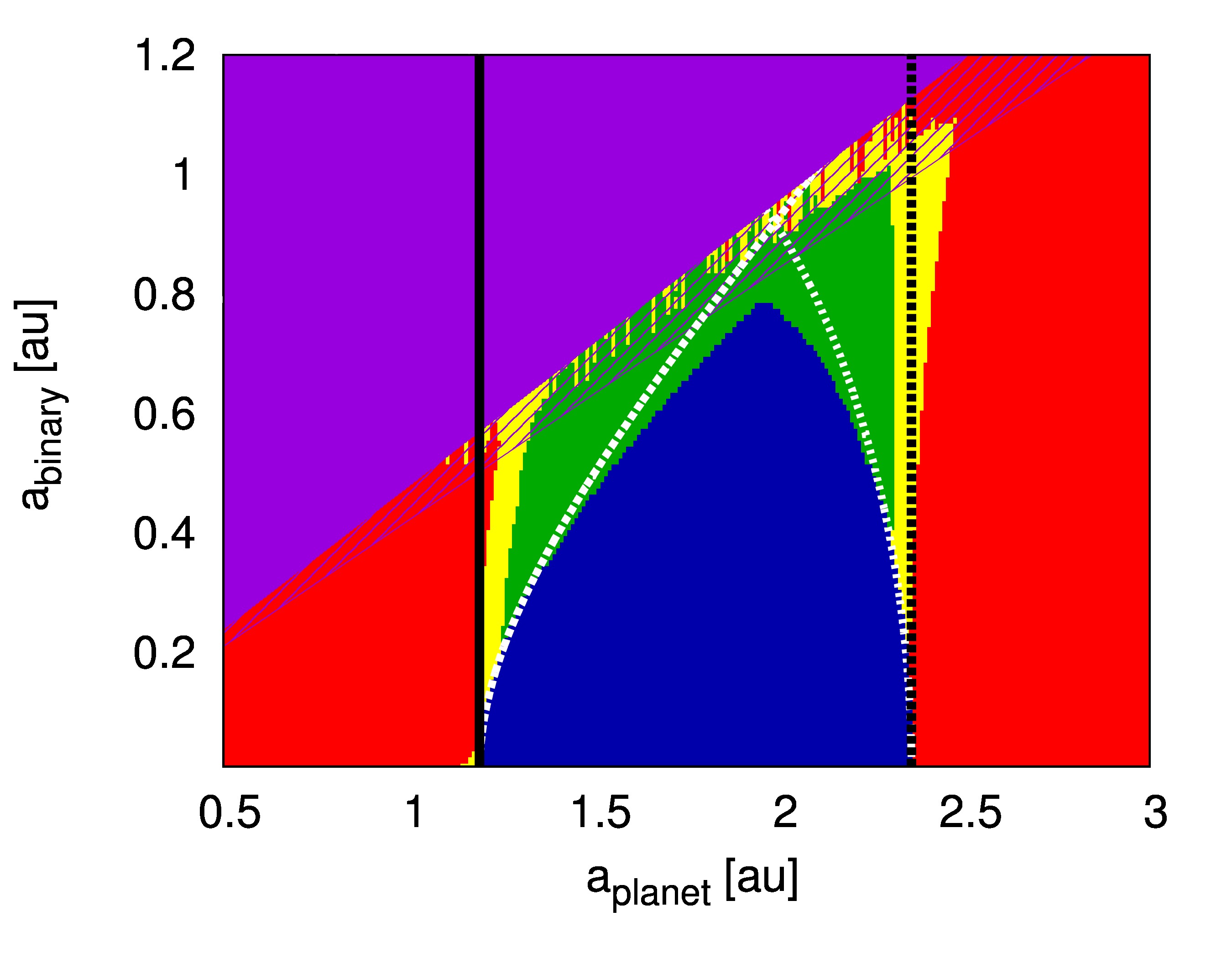}
\end{tabular}
\end{center}
\caption{PHZ, EHZ, AHZ, non-habitable and dynamically unstable regions (upper diagonal areas) are shown
for a G2V-G2V S-type system with a binary semi major axis of $a_b=10$~au (\textit{left}), 
and a set of G2V-G2V P-type systems on initially circular orbits with different semi-major axes (\textit{right}). The coloring is the same as in Figure \ref{fig1}.
The borders of the expected CHZ (to get classical estimates for P-type systems the radiation of both stars was combined in the binary's barycenter) are denoted by vertical lines.
The white curves indicate analytical estimates for the extent of the PHZ. Effective insolation values by \cite{kasting-et-al-1993} were used.
Evidently, the injected planetary eccentricities grow with the binary's eccentricity leading to a decrease in the width of the PHZs.
Even for planets around binaries on circular orbits the harsh restrictions in the PHZ become evident with growing semi-major axis of the binary star.\label{fig2}}
\end{figure}
The possible existence of terrestrial planets in and around binary star systems \citep{hatzes-2013,dumusque-et-al-2012} has rekindled the scientific community's interest in
determining Habitable Zones (HZ) in and around binary star 
systems \citep{forgan-2012,forgan-2014,kane-hinkel-2013,haghighipour-kaltenegger-2013,kaltenegger-haghighipour-2013,eggl-et-al-2012,eggl-et-al-2013, cuntz-2014, jaime-et-al-2014}.
Given the strong gravitational perturbations planets experience in such environments, significant variations in planetary orbits can be expected.
Even if planets formed on circular orbits, for instance, the interaction with the binary would force the planet's orbital eccentricity become non-zero over time. 
As was shown in e.g. \citet{eggl-et-al-2012} the 'momentary' HZ for binary star systems are solutions of 
the two simultaneous planetary insolation equations normalized with respect to effective insolation limits. In compact notation that is
\begin{equation}
\bar{S}_I \leq 1 \; \text{and} \; \bar{S}_O \geq 1,\quad \text{where} \quad \bar{S}(t)_{\{I, O\}}=\sum_{j=1}^N\frac{L_j}{4\pi S_c\{I_j,O_j\}} r_j^{-2}(t) \label{eq:2}.
\end{equation}
Here $I_j$ and $O_j$ denote the respective effective insolation limits that serve as spectral weighting factors and
 $L_j$ are the stellar luminosities; $\bar{S}_{\{I, O\}}$ is the normalized total insolation at the inner ($\bar{S}_{I}$) and outer ($\bar{S}_{O}$) border of the combined HZ, 
 and $r_j(t)$ represent the time dependent planet-to-star distances.
 For double star systems we have $N=2$. \cite{mueller-haghighipour-2014} derived a similar expression. Note, however,
 that two different sets of $\{a,b,c,d\}$ parameters, one for the inner and one for the outer HZ border 
 are required to simultaneously fulfill the inequalities in their equation~(5). 
Momentary HZs depend on the current relative positions of the stars and the planet. 
They tend to fluctuate on orbital and secular timescales.
Deriving observational constraints from time dependent 
HZs can, thus, be difficult \citep[e.g.][]{eggl-et-al-2013c}.
To include information on the variability of planetary insolation, we apply the HZ scheme defined in section 2 and 
predict planetary insolation and HZ limits analytically using recent results from perturbation theory \citep{georgakarakos-2003,georgakarakos-2005,georgakarakos-2009}. 
This allows us to model the Earth-analog's orbit and insolation evolution as a function
of the system's initial parameters \citep{eggl-et-al-2012}. 
As can be seen in Figure \ref{fig2}, PHZ and EHZ and AHZ behave quite differently for various orbital configurations of 
a terrestrial planet in a binary star system.
In all studied binary systems a clear decrease in the extent of the PHZ and EHZ can be observed for 
growing eccentricities of the binary's orbit. Of course, not only the binary's eccentricity, also its semi-major axis greatly influences the extent of permanently habitable regions.  
The right panel in Figure \ref{fig2} shows that the PHZ can be considerably smaller than the CHZ for circumbinary planets (P-type),
even if both the binary's and the planet's orbit are initially circular.
The AHZ are mostly identical with CHZs. Only in regions close to the system's stability limits 
deviations in S-type as well as P-type systems occur. 
Our HZ estimates have, furthermore, been applied to well characterized S-type binary stars close to the Solar System \citep{eggl-et-al-2013}.
It was found that most systems do not only allow for habitability in an 'average sense', they even retain zones of permanent habitability. 
\firstsection
\section{Conclusions}
The complex interplay between dynamical and radiative influences 
can turn the determination of Habitable Zones (HZ) in gravitationally interacting systems into a challenging affair.
Flexibility and adaptability should, thus, be considered important properties of any HZ calculation scheme.
We argue that a minimum dynamical model should be included in HZ calculations to ensure realistic estimates.
The proposed classification scheme (PHZ, EHZ and AHZ) retains
information on the variability of planetary insolation while providing HZ boundaries that remain valid up to stellar evolution timescales.  
It is flexible with respect to changes in atmospheric models as well as the choice of the underlying dynamical model.
An alternative to time consuming numerical insolation calculations, the analytical estimates presented in \citet{eggl-et-al-2012} can
help to shed light onto the diverse kinds of habitability occurring for terrestrial planets in gravitationally active systems.
Their application to nearby S-type binary stars has shown that most of these systems allow for HZs in spite of their strong gravitational interactions
and variable insolation conditions.
\firstsection
\section{Acknowledgments:}
The authors would like to acknowledge the support of the Austrian FWF projects S11608-N16 (sub-project of the NFN S116) and P22603-N16, the European Union Seventh
Framework Program (FP7/2007-2013) under grant agreement no. 282703, as well as the IAU Symposium no. 310 travel grant.
\firstsection

\end{document}